%% file: drsurv3.tex
\documentclass[epsf]{article}
\usepackage{graphicx}
\usepackage{epstopdf}
\usepackage{pdfpages}
\usepackage{amsmath}

\include{newcom}

\begin{document}
\baselineskip=20pt
\title{On matching-adjusted indirect comparison and calibration estimation}

\author{Jixian Wang \\
Bristol Myers Squibb}

\maketitle
\begin{abstract}
Indirect comparisons have been increasingly used  to compare data from different sources such as clinical trials and observational data in, e.g., a disease registry.  To adjust for population differences between data sources, matching-adjusted indirect comparison (MAIC) has been used in several applications including health technology assessment and drug regulatory submissions. In fact, MAIC can be considered as a special case of a range of methods known as calibration estimation in survey sampling. However, to our best knowledge, this connection has not been examined in detail.   This paper makes three contributions:  1. We examined this connection by comparing MAIC and a few commonly used calibration estimation methods, including the entropy balancing approach, which is equivalent to MAIC. 2. We considered the standard error (SE) estimation of the MAIC estimators and propose a model-independent SE estimator and examine its performance by simulation.  3. We conducted a simulation to compare these commonly used approaches to evaluate their performance in indirect comparison scenarios.     

\end{abstract}

{\bf  Key words: Calibration estimation; Empirical likelihood; Entropy balancing; Matching-Adjusted Indirect Comparison; Propensity score; Stable balancing }

\newpage

\section{Introduction}

Recently, indirect treatment comparisons have been increasingly used to compare treatments in clinical trials as well as in real world data, typically owned by multiple parties or sponsors.  A typical scenario is to compare the outcomes of treated patients, whose data are owned by one sponsor or party (e.g., a pharmaceutical company), with those of control patients, whose data are owned by another sponsor or party (e.g., a disease registry).
As the baseline characteristics of patients in the two data sources are often quite different, it is important to adjust for them in the comparison.  Although there are multiple approaches to adjust for these confounding factors, pooled individual patient data (IPD) from all parties  are often needed.  With increasing  data privacy protection, pooling IPD may not be feasible, or it may take a long time to go through a lengthy approval process.  

Some approaches can be used for confounding adjustment without pooling IPD. Among them, Matching-Adjusted Indirect Comparison (MAIC) \cite{sig10,sig12} can be used to compare clinical trial data with another data source as the target population from which only summary statistics of patient's baseline covariates and outcomes are available. MAIC uses a weighted mean outcome to compare with the mean outcome in the target population.  The weights are determined such that the weighted summary (eg, sample means) of the covariates in the trial matches that in the target population. This approach was mainly used for health technology assessment to assess treatment effects in the target population \cite{phi18}, and was recommended in a NICE Decision Support Document \cite{phi16}. It has also been used in regulatory submissions \cite{ema18}. Recently, several simulation studies for MAIC have been reported  \cite{pet18,hat20,phi20c, phi21,rem20,rem20b}. Its theoretical properties have also been examined recently \cite{zha17,che20, phi20}. 

In fact, MAIC can be considered as a special case of calibration estimation (CE) \cite{Dev92}, a powerful tool for survey sampling. CE also uses weights to balance covariates between the sample population and the target population. Furthermore, CE seeks efficient weights to reduce the variability of weighted estimators. To this end, the weights should have minimum dispersion, since equal weights are the most efficient. Using different distances to measure the dispersion leads to different estimators.  MAIC is equivalent to that maximizing the entropy of weights \cite{zha17, phi20}. Therefore, although MAIC does not aim at minimizing the dispersion, it achieves this goal anyway.  Alternative distances can be found in a seminal  CE paper \cite{Dev92}, while new ones such as the absolute difference have also been proposed.  For a recent review on CE, see Devaud and Tillé \cite{dev19}.  It is important to explore the connection between MAIC and CE, because there is rich research work on the latter. Its applications are not only found in survey sampling, but also in other areas such as missing data analysis.  Many causal inference methods such as the inverse probability weighting \cite{ros83} and its doubly robust (DR) variants also have their roots in survey sampling, in particular, CE. 

In some indirect comparisons, in addition to the treatments to be compared, the two data sources may also have a common treatment.  Therefore, if the two sources are all randomized trials, one may use the common treatment as an anchor and within-study differences for an "anchored" indirect comparison \cite{phi16}.  In some situations, this approach may eliminate the bias of unobserved baseline factors, but not that of effect modifiers. We will mainly consider unanchored comparisons to concentrate MAIC and CE. A common scenario in need of such a comparison is comparing a test treatment in a single armed trial with a control treatment in, eg, a disease registry. This scenario will be used to describe the methodology afterwards.  However, we will also discuss the use of MAIC and CE in anchored indirect comparisons and evidence generalizations.        

In this paper, we make three contributions: 1) We examined the connection between MAIC and CE, in particular three commonly used methods: entropy balancing, which is equivalent to MAIC, stable balancing weights and the empirical likelihood approach and highlight their potential use in indirect comparison;  2) We proposed a model-independent variance estimator for the MAIC estimator and examined its performance with a simulation; 3) We conducted another simulation to compare the performance of the three approaches when both the outcome model and the propensity score model were correct and when only one of them was correct. Details of implementation, especially when numerical stability is important, eg, in simulation studies are given together with R-codes, which can be easily adapted and reused.  The next section introduces indirect comparison in the framework of causal effect estimation and MAIC.  Section 3 examines the general CE and three commonly used methods.  A model-independent variance estimator is proposed in Section 4, followed by a simulation study in Section 5. Details of implementation are presented in appendices. 

\section{Causal treatment effects and MAIC}
The treatment effect to be estimated in an indirect comparison can be considered as a causal estimand.
To specify the causal estimand of interest, we take the Neyman-Rubin framework \cite{ney23,rub74}  and assume that $Y_{ij}(t)$ is the counterfactual outcome of subject $i$ in population $j$, $j=0,1$ for the external (eg, a registry) and trial populations, respectively, receiving the test treatment ($t=1$) or the control treatment ($t=0$). Denoting $E(Y_{ij}(t))=\mu_{jt}$ and $\Delta_{ij}=Y_{ij}(1)-Y_{ij}(0)$, the average treatment effect in population $j$ can be written as
    \begin{equation}
     \Delta_j=\mu_{j1}-\mu_{j0} = E(\Delta_{ij})
     \label{delta}
    \end{equation}
where the expectation is taken over the population $j$. Our main focus is on estimating $\Delta_0=\mu_{01}-\mu_{00}$ in an unanchored indirect comparison when the target population is treated with $t=0$ and the trial population is treated with $t=1$ only, ie, $t=j$. Specifically, we assume that IPD $Y_{i1},T_{i1},\bX_{i1}, i=1,...,n$ are available in the trial. However, only summary statistics of covariates   in the target population are accessible.   A different, but symmetric task is borrowing controls from a disease registry for a clinical trial, with IPD available in the external source, but cannot be pooled with the trial IPD. Nevertheless,  the analyst in the registry will have access to her own data and summary statistics of the trial data. Although the goal is to estimate $\Delta_1$: treatment effect in the trial population, the MAIC and other CE approaches below apply exactly as in the task we are focusing.  

Let $\bar{\bX}_0$ be the mean of $\bX_{i0}$ in the target population,   the MAIC approach aims at balancing the summary statistics of $\bX_{ij}$ between the target and the trial populations by weighting the $n$ trial subjects such that  $\bar{\bX}_{0}= \sum_{i=1}^n w_i \bX_{i1}$.  This can be achieved by determining parameters $\bgamma$  so that the constraints are satisfied by weights
\begin{equation}
    w_i=\exp(\bgamma^T \bX_{i1})/\sum_{i=1}^{n} \exp(\bgamma^T \bX_{i1}).
\end{equation} To this end, one needs to find $\bgamma$ with a numerical algorithm\cite{sig10}.  After obtaining the weights, $\Delta_0$ can be estimated by the MAIC estimator: 
\begin{equation}
    \hat \Delta_{0}^c=\sum_{i=1}^n w_i Y_{i1} -\bar{Y}_0
\end{equation}
where $\bar{Y}_0$ is the sample mean outcome in population 0 as an estimator for  $\mu_{00}$. 

To see how it works, suppose that the outcome model (Y-model henceforth) is
\begin{equation}
 Y_{ij}(t)=\bbeta_t^T \bX_{ij}+\epsi_{ij}   
\label{outcome}
\end{equation}
where $\bX_{ij}$ includes an intercept and $\epsi_{ij}$ is an independent zero-mean error term. Therefore, the treatment effect in the target population is $\Delta_{0}=E(\bX_{i0})^T(\bbeta_1-\bbeta_0)$.  Then,
\begin{equation}
    E(\hat \Delta_{0}^c)=\sum_{i=1}^n E(w_i \bX_{i1}^T \bbeta_1)-E(\bar Y_0)= E(\bar{\bX}_0)^T (\bbeta_1-\bbeta_0)=E(\bX_{i0})^T (\bbeta_1-\bbeta_0).
\end{equation}
where the second equation is due to the constraints $\bar{\bX}_{0}= \sum_{i=1}^n w_i \bX_{i0}$. Therefore, MAIC does eliminate the bias due to the difference in $\bX_{ij}$ when the Y-model is a linear function of $\bX_{ij}$ and $\hat \Delta_0^c$ estimates the treatment effect in the right population.   As $\mu_{00}$ can be well estimated by the sample mean $\bar Y_0$,  we will concentrate on the estimation of $\mu_{01}=E_0(Y_{i0}(1))$, that is, the mean response of the active treatment in the target population $j=0$.  The estimators for $\mu_{01}$ we are considering all have the form of
\begin{equation}
    \hat \mu_{01}=\sum_{i=1}^n w_i Y_{i1} 
    \label{weiest}
\end{equation}
with $w_i$s determined in different ways, as shown in the next section.

MAIC and CE approaches can also be used for two similar tasks: 1) to generalize evidence from a randomized controlled trial (RCT) ($j=1$) to the target population ($j=0$); 2) to compare treatment 0 and 1 when there is a common treatment, say, 2, in two  RCTs.  That is, one RCT has treatments 0 and 2, and the other has treatments 1 and 2.   
In the first task,  $T_{i1}$ is randomized so that one can estimate $\Delta_1$ based on the trial data without any adjustment. Our goal is to estimate $\Delta_0$, based on $\bar \bX_0$ only from population $0$.   With $T_{i1}$ randomized, one can estimate $\Delta_1$ by
\begin{equation}
\sum_{i=1}^n Y_{i1} (\frac{T_{i1}}{n_1}-\frac{1-T_{i1}}{n_0})
\end{equation}
where $n_t$ is the number of subjects treated with $t$. With $w_i$ satisfying $\bar{\bX}_{0}= \sum_i^n w_i \bX_{i1}$, we can use
\begin{equation}
\hat \Delta_0^g=\sum_{i=1}^n w_i Y_{i1} (\frac{T_{i1}}{n_1}-\frac{1-T_{i1}}{n_0})
\end{equation}
to estimate $\Delta_0=E(\bX_{0})^T (\bbeta_1-\bbeta_0)$, because 
\begin{equation}
E(\hat \Delta_0^g)=E(\sum_{i=1}^n w_i \bX_{i1}^T (\bbeta_1-\bbeta_0))=E(\bX_{0})^T (\bbeta_1-\bbeta_0).
\end{equation}
Therefore, the key step of this task is also to find $w_i$ to balance the covariates between the two populations.  For the other task with $\Delta_0$ as the estimand, one can construct an estimator as
\begin{equation}
\hat \Delta_0^a=\sum_{i=1}^n w_i Y_{i1} (\frac{T_{i1}}{n_1}-\frac{1-T_{i1}}{n_2})-(\hat \mu_{00} -\hat \mu_{02})
\end{equation}
where $T_{i1}=1$ for treatment $1$ and $T_{i1}=0$ for treatment $2$ in population 1,  $\hat \mu_{00}$ and $\hat \mu_{02}$  are the mean responses under treatment 0 and 2 in population 0, respectively.  This is known as an anchored comparison by the common treatment $2$.  Although both tasks use the MAIC weights in the same way as the unanchored indirect comparison, their assumptions on the types of confounders need careful consideration. 

Model (\ref{outcome}) considers each covariate in $\bX_{ij}$ as an effect modifier, which modifies treatment effects, unless the corresponding element in $\bbeta_1-\bbeta_0$ equals to zero. Based on the model, we can  investigate the impact of imbalance  measured by $\bD= \sum_{i=1}^n w_i \bX_{i1}-\bar{\bX}_{0}$.  This imbalance may be due to either unobserved covariates or the difficulty of exactly balancing some covariates, even when they are included in the $\bar{\bX}_{0}= \sum_i^n w_i \bX_{i1}$. For the unanchored indirect comparison, we can write 
\begin{equation}
    E(\hat \Delta_0^c)= E(\bar{\bX}_0)^T (\bbeta_1-\bbeta_0) +\bD^T \bbeta_0.
    \label{bias}
\end{equation}hence the bias term is $\bD^T \bbeta_0$.  Therefore, any imbalanced element in $\bX_{ij}$, regardless of effect modifier or not, contributes to the bias in the estimate.   In contrast, for generalization, 
\begin{equation}
E(\hat \Delta_0^g)=E(\bX_{0})^T (\bbeta_1-\bbeta_0) +\bD^T (\bbeta_1-\bbeta_0).
\end{equation}Therefore, only effect modifiers (those with non-zero elements in $\bbeta_1-\bbeta_0$) contribute to the bias $\bD^T (\bbeta_1-\bbeta_0)$.  The same holds for the anchored indirect comparison estimator $\hat \Delta_0^a$.  Since effects of baseline factors in both studies have been eliminated in the weighted mean difference and $\hat \mu_{00} -\hat \mu_{02}$, only effect modifiers cause biases.  Therefore, in both generalization and anchored indirect comparison, imbalanced baseline factors do not invalidate the estimators.  Consequently, Ref \cite{phi16} recommended including effect modifiers only in the constraints for anchored indirect comparisons.

Although often people tend to balance as many potential confounding factors as possible, this may lead to extreme weights and a highly variable estimator.   Calibration estimation, as described in the next section, provides multiple tools to achieve a good compromise between covariate balancing and less variable weights.  
 Since the key step for all above tasks is to find $w_i$ such that $\bar{\bX}_{0}= \sum_{i=1}^n w_i \bX_{i1}$, in the next section, we will write $\bX_{i1}$ as $\bX_{i}$, $\bbeta_1$ as $\bbeta$, $\Delta_0$ as $\Delta$ and $\mu_{01}$ as $\mu_1$  for simplicity.

\section{ Connection with calibration estimation}
A general CE  approach \cite{Dev92}, adapted for our context, is to find weights $w_i, i=1,..., n$ as the solution to 
\begin{align}
    & \min_{w_i} \sum_{i=1}^n D(w_i,1/n) \nn\\
    \mbox{subject to:} & \sum_{i=1}^n w_i \bX_i=\bar{\bX}_0 
    \label{gence}
\end{align}
where $D(w_i,1/n)$ is a distance between $w_i$ and the uniform  weight  $1/n$.  Then the weighted mean (\ref{weiest}) is used to estimate $\mu_1$.  The general idea is to minimize the distance between $w_i$ and the uniform weight $1/n$, which is the most efficient. Among several common choices for the distance is the entropy $w_i \log(w_i)$, which yields entropy balancing weights \cite{hai11} as the solution to 
\begin{align}
    & \min_{w_i} \sum_{i=1}^n w_i \log(w_i) \nn\\
    \mbox{subject to:} &\sum_{i=1}^n w_i \bX_i=\bar{\bX}_0,\nn\\
      &\sum_{i=1}^n w_i=1.
      \label{ebal}
\end{align}
In fact, the entropy balancing weights are the same as the MAIC weights, although the two methods look rather different \cite{che20,phi20}. They are also equivalent to the Case 2 weights in \cite{Dev92} with constraint $\sum_{i=1}^n w_i=1$. Another popular choice is the quadratic distance $D(w,d)=(w-1/n)^2$ \cite{Dev92}, which leads to a closed formula for $w_i$.  A recent paper examined this approach in the context of indirect comparison and showed that a larger effective sample size, defined as $ESS=(\sum_{i=1}^n w_i)/\sum_{i=1}^n w_i^2$, than that of MAIC can be obtained \cite{jac21}.  This result is not surprising at all, as $\sum_{i=1}^n (w_i-1/n)^2$ is a monotone decreasing function of ESS, hence the quadratic distance based calibration  maximizes ESS under the given constraints. The above may serve as an example for considering alternatives to MAIC in the CE framework.
The major issue of this approach is that some $w_i$s might have negative values. However, one may add constraint $w_i \ge 0$ in (\ref{gence}).  Zubizarreta \cite{zub15} proposed stable balancing weights by forcing $w_i \ge 0$ but relaxing the constraints in (\ref{gence}) to
$|\sum_{i=1}^n w_i \bX_i-\bar{\bX}_0| < \bd $ to allow for slight imbalance, controlled by a vector $\bd$ to make the weights more stable. Since $\bd$ controls $\bD$ in (\ref{bias}),  one can choose $\bd$ based on the impact of each element in $\bX_i$ on the outcome.    

An intuitive approach to estimating $\mu_1$  is to fit a model $E(Y_i)=\bX_i^T \bbeta$ with the trial data, then use $\hat \mu_1^{reg}=\bar X_0 \hat \bbeta$ where $\hat \bbeta$  are parameter estimates in the fitted model.  This approach is known as generalized regression estimation in CE, which is the same or very similar, depending on the way of implementation, to the Simulated Treatment Comparison (STC) approach in Ref \cite{phi16}.  One interesting connection to the weighting approach is that, under some technical conditions, a weighted CE estimator is (asymptotically) equivalent to $\hat \mu_1^{reg}$, although no model is fitted explicitly \cite{Dev92}. 

Another important CE approach is based on the empirical likelihood \cite{che93, rao06} that finds $w_i$s to maximise
\begin{equation}
    \sum_{i=1}^n \log(w_i),
\end{equation}
subject to the same constrains as in (\ref{gence}) and use (\ref{weiest}) as the estimator.  This is equivalent to solving (\ref{gence}) with $D(w_i,1/n)=-\log(w_i)$.      This approach produces nonnegative weights, but some weights may be extreme \cite{che93,rao06}.  The calculation of weights also requires solving nonlinear equations numerically, as detailed in the appendix. 

Although apparently MAIC is valid only when the Y-model is linear with respect to $\bX_i$, due to its equivalence to the entropy balancing approach, it is also valid when the logit of propensity score \cite{ros83}  $p_i=P(i \in \mbox{population 1}|\bX_i)$ is a linear function of $\bX_i$, that is, $p_i=1/(1+\exp(-\balpha^T\bX_i))$ with parameters $\balpha$ (P-model henceforth).  More specifically, $w_i$ is the inverse odds of $p_i$ if this condition holds \cite{zha17}.  The entropy balancing estimator (\ref{weiest}) with weights determined by (\ref{ebal}),  is doubly robust (DR) in the sense that the estimator is consistent when either the Y-model or the logit of the P-model is a linear function of $\bX_i$.  This fact provides reinsurance of using MAIC in situations with model uncertainty.
In fact, under some technical conditions, a wide range of CE estimators derived from (\ref{gence}) are DR  in the sense that the distance measure specifies an underline PS model and the estimator is valid if the true PS model has the same form \cite{wan20}.  
For example, the weights that minimize $(w_i-1/n)^2$ is valid if the true propensity score has the form of $p_i=(1/n-\balpha^T \bX_i)^{-1}$. Nevertheless, the common construction of $\bX_i$, eg, from multi-normal distributions with different means leads to $p_i=(1+\exp(-\balpha^T \bX_i))^{-1}$, which makes MAIC, but not that minimising $(w_i-1/n)^2$, valid regardless of the Y-model,  as indicated by the simulation results in Section 5.    

\section{Variance estimation for MAIC esitmator}
Statistical inference for $\Delta_0$ based on any weighted estimator often depends on the estimation of its SE, or equivalently, its variance.  Although the MAIC estimator for $\Delta_0$ is doubly robust, its variance is not easy to estimate. First, with only summary statistics for $Y_i,\bX_i$ from the target population, it is impossible to obtain accurate SE estimates without strong assumptions.  We will consider two situations.  In the first,  the target population is fixed, hence we only need to consider the variance of $\hat \mu_1=\sum_{i=1}^n w_i Y_i$.  Although seems unreasonable, it may be (approximate) valid when the target data source is much larger than the trial sample size, or it approximately represents the target population. For example,  a national disease registry may be considered to represent the majority of patients in a country, even the number of patients is not very large.  In this case, the treatment effect of interest can be considered as that in the registry, had the test treatment been applied to its patients.

Even with this simplification, estimating the variance of $\hat \mu_1=\sum_{i=1}^n w_i Y_i$ may still not be easy  in the general situation when only one of the Y- and P-models is correct.  The original sandwich estimator proposed in \cite{sig10}  seemed to be
\begin{equation}
  \hat V_0=\sum_{i=1} w_i^2 (Y_i- \hat \mu_1)^2    
\end{equation}
Chen et al. \cite{che20} derived a more accurate variance, assuming a correct P-model. The calculation is rather complex, even with some suggested simplifications.  

$\hat V_0$ can be rather conservative, as the fact that the weights are to balance the covariates is not taken into account. For example, when $Y_i$ follows a linear model and $V(Y_i|\bX_i)=\sigma^2$, one can write
\begin{align}
    V(\sum_{i=1}^n w_i Y_i) &= E(\sum_{i=1}^n V(w_i Y_i|\bX_i)) + V(E(\sum_{i=1}^n w_i Y_i|\bX_i))\\
     &= E(\sum_{i=1}^n w_i^2 V(Y_i|\bX_i)) +V(\bbeta^T \bar{\bX}_0)\\
     & =\sum_{i=1}^n w_i^2 \sigma^2
\end{align}
where the second equation used the balancing condition and the third one the assumption that $\bar{\bX}_0$ is fixed.  This shows that, in this situation, the variability in $\hat \mu_1$ due to $\bX_i$ is eliminated by weighting, but is included in $\hat V_0$.    In fact, this property holds for calibration estimators with different distance functions in \cite{Dev92}.   A common variance estimator in survey sampling is 
\begin{equation}
 \hat V_{ss} =\sum_{i=1}^n w_i^2 (Y_i-\hat m(\bX_i))
\end{equation}
where $\hat m(\bX_i)$ is a fitted Y-model.  This estimator does not have the issue of being conservative.  But it depends on the fitted outcome model.

A model-independent estimator for $\var(\hat \mu_1)$ can be constructed by considering $\hat \mu_1$ as a result of the two-step generalized method of moments approach, in which the asymptotic variance of the estimator for the parameter of interest is well established. The reader is referred to Newey  \cite{new94}. For this purpose, we define two estimating equations, one for the estimation of $\bgamma$ and another for $\mu_1$:
\begin{align}
    S_1(\bgamma) &= \sum_{i=1}^n w_i (\bX_i - \bar{\bX}_0)\\
    S_2(\mu_1,\bgamma) &=  \sum_{i=1}^n w_i (Y_i- \mu_1),
\end{align}
and their individual components as
$S_{1i}=w_i (\bX_i - \bar{\bX}_0)$ and $S_{2i}=w_i (Y_i - \mu_1)$, with their dependency on parameters suppressed for simplicity.  $\hat \mu_1$ and $\hat \bgamma$ are solutions to these equations.  A sandwich variance estimator for $\hat \mu_1$ is (Theorem 6.1 in Ref.  \cite{new94})
\begin{equation}
  \hat V_{2s}=  (\sum_{i=1}^n \frac{\partial S_{2i}}{\partial \mu_1})^{-1} \sum_{i=1}^n S_i^2
    (\sum_{i=1}^n \frac{\partial S_{2i}}{\partial \mu_1})^{-1}=\sum_{i=1}^n S_i^2,
\end{equation}
where the second equation is due to $\partial S_{2i}/ \partial \mu_1=-w_i$ and $\sum_{i=1}^n w_i=1$, and
\begin{equation}
    S_i= S_{2i} - \sum_{i=1}^n \frac{\partial S_{2i}}{\partial \bgamma}
    (\sum_{i=1}^n \frac{\partial S_{1i}}{\partial \bgamma})^{-1} S_{1i} 
\end{equation}
with 
\begin{align}
    \partial S_{2i}/ \partial \bgamma &= w_i \bX_i  (Y_i-\mu_1)\\
    \partial S_{1i}/ \partial \bgamma &= w_i \bX_i (\bX_i-\bar{\bX}_0)^T
\end{align}
Note that this calculation does not involve any fitted model. It can also be considered as a simplified version of Seaman et al. \cite{sea18} (Section 3.3) where a  Y-model was used to form a doubly robust estimator.  

Like other variance estimators which depend on asymptotic properties, this estimator may not be accurate when the sample size is small.  A simulation to examine its accuracy, compared with $\hat V_0$, is presented in the next section.  An alternative is bootstrap, which can be used either to estimate the variance with a moderate number of bootstrap runs, or to calculate confidence intervals with many runs  (non-parametric bootstrap).  Simulation evaluation for the nonparametric bootstrap approach is difficult due to the need of intensive computation and also the instability of balancing algorithms for some bootstrapped datasets \cite{phi19}.

When considering $\bar{Y}_0$ and $\bar{\bX}_0$ as fixed is not acceptable, one needs additional information and assumptions to estimate $\var(\hat \Delta^c)$.
When $\var(Y_i)=\sigma_0^2$ in the target data is reported, one may use
\begin{equation}
    \var(\sum_{i=1}^n w_i Y_i)+\sigma_0^2/n_0
\end{equation}
where the first one can be estimated as above, $n_0$ is the sample size of the target source. This is likely a conservative estimate. For example, when $\bX_i$ is an indicator of a categorical factor, $\hat \Delta^c$ is a weighted mean of stratified differences, in which the within-stratum variance may be much smaller than the marginal variance $\sigma_0^2$.  However, it is difficult to improve it without further assumptions, eg, homogeneous variance conditional on $\bX_i$ across populations.  

\section{A simulation study}
One purpose of this simulation study is to evaluate the proposed SE estimator, compared with the one originally proposed for MAIC and that based on bootstrap.  As discussed before, we concentrate on the estimation of $\mu_1$ rather than $\Delta$. We examine the bias and variance estimators in three scenarios: both models are correct; the P-model is correct but the Y-model is not (ie, it is not a linear function of $\bX_i$); The Y-model is correct but the P-model is not (ie, the log-odds of $p_i$ is not a linear function of $\bX_i$).  
The simulation settings are as follows 
\begin{enumerate}
    \item $\bX_i \sim N(\bm,\bI_p)$ with means $\bm$ and $\bI_p$ is a $p$-identity matrix.   $\bar \bX_0$ is the mean of $n_0$ samples  from $\bX_{i0} \sim N(0,\bI_p)$ with $n_0=2000$ and is considered fixed. 
    \item In the correct Y-model scenario $Y_i=\bbeta^T \bX_i+\epsi_i$ with $\var(\epsilon_i)=1$ and $\bbeta=\beta \bone_p$ ($\bone_p$ is a vector of $p$ 1s).  
    \item In the incorrect Y-model scenario $Y_i=I(\bbeta^T \bX_i+\epsi_i>0)$, where $I(a)=1$ if $a$ is true and $I(a)=0$ otherwise.
    \item The incorrect P-model was created by $\bX_i=\exp(0.5 \bZ_i)$ where  $\bZ_i  \sim N(\bm,\bI_p)$. 
    \item One simulation had a base scenario with $n_1=500$, $p=3$, $\bm=b \bone_p$,  and $\beta=0.3$. Other scenarios were created with varying values of these parameters.
    \item Another simulation used a base scenario as above, except that the first 2 components of $\bm$ were set at 0.5, the others at 0.25. Other parameters were varying.  
\end{enumerate}
For each scenario, 2000 simulation runs were generated. Appendix 1 gives details about the implementation of these weight calculation algorithms.  A sample R-code can be found in Appendix 2. For each simulation, 50 bootstrap runs were used to estimate the bootstrapped variance. A non-parametric bootstrap was also attempted to construct 95\% confidence intervals (CI), but resulted in a large number of runs with numerical issues, hence the results will not be presented here.    

The simulation results based on the first base scenario with a variety of parameter settings and different model assumptions are summarized in Table 1. Summary statistics include the bias of MAIC estimator and its SEs based on $\hat V_{2s}$ (2S), the bootstrapped variance (Boot) $ \hat V_0$ (MAIC) and the empirical SE (Emp.) calculated from the simulation runs.  Also presented are the coverage of the 95\% CI based on each estimated SE and the normal approximation.  As a reference scale, the bias of unadjusted estimates is also presented. The coverage of 95\% CI based on the MAIC SE is not presented, as it was almost always close to 100\%, due to its very conservative SE estimates.  In general, even with Y- and P-models correctly specified, the coverages based on $\hat V_{2s}$ and the bootstrapped variance were less than the nominal level, with the worst results occurred when $n_1=100$ and when $p=7$. This is partly due to underestimated SEs, partly due to non-normality of $\hat \mu_1$.  For example, in the second row the SE of Boot is 0.051, the same as the empirical one, but the coverage is 0.93. 
Increasing $n_1$ to 1000 led to almost correct coverage and SE estimates by both Boot and 2S.  The same also happened when the mean differences $b$ between $\bX_{i}$ and $\bar \bX_0$ were reduced to 0.25. 

An incorrect Y-model could make the issues of the 95\% CI coverage and the bias in the SE even worse, with the worst also occurring when $p=7$ and when $n_1=100$. The biases in $\hat \mu_1$ increased slightly, but the size was no more than 0.02, compared with the maximum unadjusted bias around 0.4.   This result suggests that although the MAIC estimator is DR, the 2S SE estimator seemed less robust to Y-model misspecification. Also affected were the bootstrapped 95\% CI and SE estimates, although they were model-independent.
Nevertheless, almost correct coverage and SE estimates with both Boot and 2S SEs were also achieved when $b=0.25$.

Surprisingly, the results with the incorrect P-model were better than those with both models correct, although the similar patterns were found. However, due to the P-model settings, the unadjusted biases were much less than those when both models were correct. As the Y-model was correctly specified, the smaller unadjusted bias must be due to smaller differences between the means of $\bX_{i}$ and $\bar \bX_0$.  These results showed that the SE estimators were more robust to misspecified P-model, as long as the Y-model is correct. 

To further explore the impact of $\bm$, a simulation for the second set of scenarios based a varying patterns of $\bm$ was also conducted.  The results are presented in Table 2. In general, very similar patterns across different settings were found.  However, both coverage and SE estimates were less biased than in the corresponding scenarios in Table 1, especially when $p=7$. As there were only two large elements in $\bm$, it is easier to match $\bX_0$ by $\sum_{i=1}^n w_i \bX_i$ than in the first set of simulations.  Consequently, the SE estimators and 95\% CIs performed better.

\begin{table}[ht]
\caption{Summary of simulations to  examine the bias, the coverage of 95\% CI, SEs based on the variance estimators (2S:$\hat V_{2s}$,  Boot:bootstrapped variance and MAIC:$\hat V_0$), compared with the empirical SE (Emp.), when both models are correct and when only the Y-model or the P-model is incorrect.    \label{Tab1}}
\centering
\begin{tabular}{rrrrrrrrrrrr}
  \hline
  & & & &  \multicolumn{2}{c}{Bias} & \multicolumn{2}{c}{CI coverage} & \multicolumn{4}{c}{SE}\\
 $n_1$ & $\beta$ & $b$ & $p$ & Unadj. & MAIC & 2S& Boot &  2S& Boot & MAIC  & Emp. \\ 
  \hline
  \multicolumn{12}{c}{Both models are correct}\\
  \hline
100 & 0.30 & 0.50 & 3 & 0.460 & -0.001 & 0.898 & 0.919 & 0.066 & 0.073 & 0.112 & 0.074 \\ 
200 & 0.30 & 0.50 & 3 & 0.462 & -0.000 & 0.926 & 0.933 & 0.049 & 0.051 & 0.083 & 0.051 \\ 
500 & 0.30 & 0.50 & 3 & 0.460 & 0.000 & 0.929 & 0.922 & 0.032 & 0.033 & 0.054 & 0.034 \\ 
1000 & 0.30 & 0.50 & 3 & 0.461 & -0.000 & 0.939 & 0.941 & 0.023 & 0.023 & 0.039 & 0.024 \\ 
200 & 0.30 & 0.25 & 3 & 0.236 & 0.000 & 0.942 & 0.939 & 0.039 & 0.039 & 0.059 & 0.039 \\ 
500 & 0.30 & 0.75 & 3 & 0.685 & 0.000 & 0.898 & 0.907 & 0.046 & 0.048 & 0.083 & 0.052 \\ 
500 & 0.30 & 0.50 & 5 & 0.763 & -0.000 & 0.912 & 0.914 & 0.038 & 0.040 & 0.082 & 0.043 \\ 
500 & 0.30 & 0.50 & 7 & 1.056 & -0.001 & 0.880 & 0.905 & 0.044 & 0.048 & 0.112 & 0.053 \\ 
  \hline
 \multicolumn{12}{c}{Y-model is incorrect} \\
 \hline 
 100 & 0.30 & 0.50 & 3 & 0.314 & -0.011 & 0.884 & 0.907 & 0.037 & 0.042 & 0.073 & 0.042 \\ 
 200 & 0.30 & 0.50 & 3 & 0.309 & -0.007 & 0.903 & 0.908 & 0.027 & 0.029 & 0.051 & 0.031 \\ 
 500 & 0.30 & 0.50 & 3 & 0.307 & -0.004 & 0.917 & 0.914 & 0.018 & 0.019 & 0.032 & 0.020 \\ 
 1000 & 0.30 & 0.50 & 3 & 0.307 & -0.003 & 0.920 & 0.921 & 0.013 & 0.014 & 0.023 & 0.014 \\ 
 200 & 0.30 & 0.25 & 3 & 0.166 & -0.002 & 0.950 & 0.945 & 0.023 & 0.023 & 0.039 & 0.023 \\ 
 500 & 0.30 & 0.75 & 3 & 0.409 & -0.010 & 0.861 & 0.876 & 0.024 & 0.026 & 0.049 & 0.028 \\ 
 500 & 0.30 & 0.50 & 5 & 0.367 & -0.005 & 0.897 & 0.907 & 0.021 & 0.022 & 0.041 & 0.024 \\ 
 500 & 0.30 & 0.50 & 7 & 0.409 & -0.020 & 0.774 & 0.816 & 0.023 & 0.026 & 0.050 & 0.028 \\ 
      \hline
 \multicolumn{12}{c}{P-model is incorrect} \\
 \hline 
 100 & 0.30 & 0.50 & 3 & 0.287 & 0.001 & 0.924 & 0.932 & 0.063 & 0.065 & 0.075 & 0.066 \\ 
 200 & 0.30 & 0.50 & 3 & 0.287 & 0.001 & 0.940 & 0.935 & 0.046 & 0.046 & 0.053 & 0.047 \\ 
 500 & 0.30 & 0.50 & 3 & 0.286 & 0.000 & 0.949 & 0.944 & 0.029 & 0.029 & 0.033 & 0.030 \\ 
 1000 & 0.30 & 0.50 & 3 & 0.287 & 0.000 & 0.951 & 0.946 & 0.021 & 0.021 & 0.024 & 0.021 \\ 
 200 & 0.30 & 0.25 & 3 & 0.133 & 0.001 & 0.949 & 0.945 & 0.038 & 0.038 & 0.043 & 0.038 \\ 
 500 & 0.30 & 0.75 & 3 & 0.460 & 0.001 & 0.944 & 0.936 & 0.040 & 0.040 & 0.046 & 0.041 \\ 
 500 & 0.30 & 0.50 & 5 & 0.472 & -0.001 & 0.942 & 0.938 & 0.034 & 0.035 & 0.044 & 0.035 \\ 
 500 & 0.30 & 0.50 & 7 & 0.672 & 0.001 & 0.925 & 0.932 & 0.040 & 0.042 & 0.059 & 0.043 \\    
 \hline
\end{tabular}
\end{table}

\begin{table}[ht]
\caption{Summary of simulations to  examine the bias, the coverage of 95\% CI, SEs based on the variance estimators (2S:$\hat V_{2s}$,  Boot:bootstrapped variance and MAIC:$\hat V_0$), compared with the empirical SE (Emp.), when both models were correct and when only the Y-model or the P-model was incorrect.  The first two elements in $\bm$ were 0.5 and the rest were 0.25.     \label{Tab2}}
\centering
\begin{tabular}{rrrrrrrrrrr}
  \hline
  & & &  \multicolumn{2}{c}{Bias} & \multicolumn{2}{c}{CI coverage} & \multicolumn{4}{c}{SE}\\
 $n_1$ & $\beta$ & $p$ & Unadj. & MAIC & 2S& Boot &  2S& Boot & MAIC  & Emp. \\ 
  \hline
  \multicolumn{11}{c}{Both models are correct}\\
  \hline
 100 & 0.30 & 3 & 0.385 & -0.001 & 0.915 & 0.927 & 0.062 & 0.067 & 0.102 & 0.068 \\ 
200 & 0.30 & 3 & 0.387 & -0.000 & 0.938 & 0.938 & 0.045 & 0.047 & 0.075 & 0.047 \\ 
500 & 0.30 & 3 & 0.385 & 0.000 & 0.939 & 0.930 & 0.029 & 0.030 & 0.048 & 0.031 \\ 
1000 & 0.30 & 3 & 0.386 & -0.000 & 0.943 & 0.938 & 0.021 & 0.021 & 0.034 & 0.021 \\ 
500 & 0.30 & 5 & 0.537 & -0.000 & 0.937 & 0.934 & 0.031 & 0.031 & 0.061 & 0.033 \\ 
500 & 0.30 & 7 & 0.681 & -0.001 & 0.927 & 0.926 & 0.032 & 0.033 & 0.074 & 0.035 \\ 
  \hline
 \multicolumn{11}{c}{Y-model is incorrect} \\
 \hline 
100 & 0.30 & 3 & 0.269 & -0.008 & 0.922 & 0.931 & 0.036 & 0.039 & 0.067 & 0.038 \\ 
200 & 0.30 & 3 & 0.265 & -0.005 & 0.921 & 0.926 & 0.026 & 0.027 & 0.047 & 0.028 \\ 
500 & 0.30 & 3 & 0.265 & -0.004 & 0.921 & 0.918 & 0.017 & 0.017 & 0.030 & 0.018 \\ 
 1000 & 0.30 & 3 & 0.264 & -0.003 & 0.923 & 0.920 & 0.012 & 0.012 & 0.021 & 0.013 \\ 
 500 & 0.30 & 5 & 0.279 & -0.002 & 0.916 & 0.922 & 0.018 & 0.018 & 0.032 & 0.019 \\ 
 500 & 0.30 & 7 & 0.299 & -0.014 & 0.844 & 0.852 & 0.018 & 0.019 & 0.033 & 0.020 \\ 
      \hline
 \multicolumn{11}{c}{P-model is incorrect} \\
 \hline 
  100 & 0.30 & 3 & 0.235 & 0.001 & 0.930 & 0.932 & 0.060 & 0.061 & 0.070 & 0.062 \\ 
 200 & 0.30 & 3 & 0.235 & 0.001 & 0.943 & 0.938 & 0.043 & 0.043 & 0.050 & 0.044 \\ 
 500 & 0.30 & 3 & 0.235 & 0.000 & 0.949 & 0.947 & 0.027 & 0.028 & 0.031 & 0.028 \\ 
 1000 & 0.30 & 3 & 0.236 & 0.000 & 0.952 & 0.944 & 0.019 & 0.019 & 0.022 & 0.019 \\ 
 500 & 0.30 & 5 & 0.318 & -0.000 & 0.950 & 0.945 & 0.029 & 0.029 & 0.036 & 0.029 \\ 
 500 & 0.30 & 7 & 0.416 & 0.000 & 0.946 & 0.944 & 0.030 & 0.030 & 0.041 & 0.030 \\ 
 \hline
\end{tabular}
\end{table}

The second goal of this simulation study was to compare the performance of the three estimators in the form of (\ref{weiest}) using: MAIC weights;  stable balancing weights (SBW) with $\bd=0.005 \bone_p$, and empirical likelihood weights (ELW).  Many simulation studies for CE estimator comparison in survey sampling research concentrated on its own scenarios, typically with very large sample sizes, while ours focused on indirect comparison.     The simulation settings are similar to those in the first part, with some modifications.  The sample size was fixed at $n_1=200$ to concentrate on other factors.   Again three scenarios were generated: both models are correct (denoted as Y:R,P:R); only Y-model is correct (Y:R, P:W) and only P-model is correct (Y:W, P:R),  with 1000 simulation runs each. The error distributions of the three estimators are shown in the box plots in Figure 1.  The MAIC estimator had similar performance to the SBW estimator in most cases, while the ELW estimator showed a much larger variability than the others.  All estimators were unbiased when the Y-model is correctly specified, with and without the correct P-model. The MAIC estimator had no bias when the Y-model was wrong, since the P-model is correct. However, the SBW and ELW estimators were biased, since the P-model was also wrong for them (Section 3).  In the worst case with $b=0.5$ and $p=7$, both SBW and ELW  estimators showed moderate biases in opposite directions.  
\begin{figure}
    \centering
    \includegraphics[width=160mm]{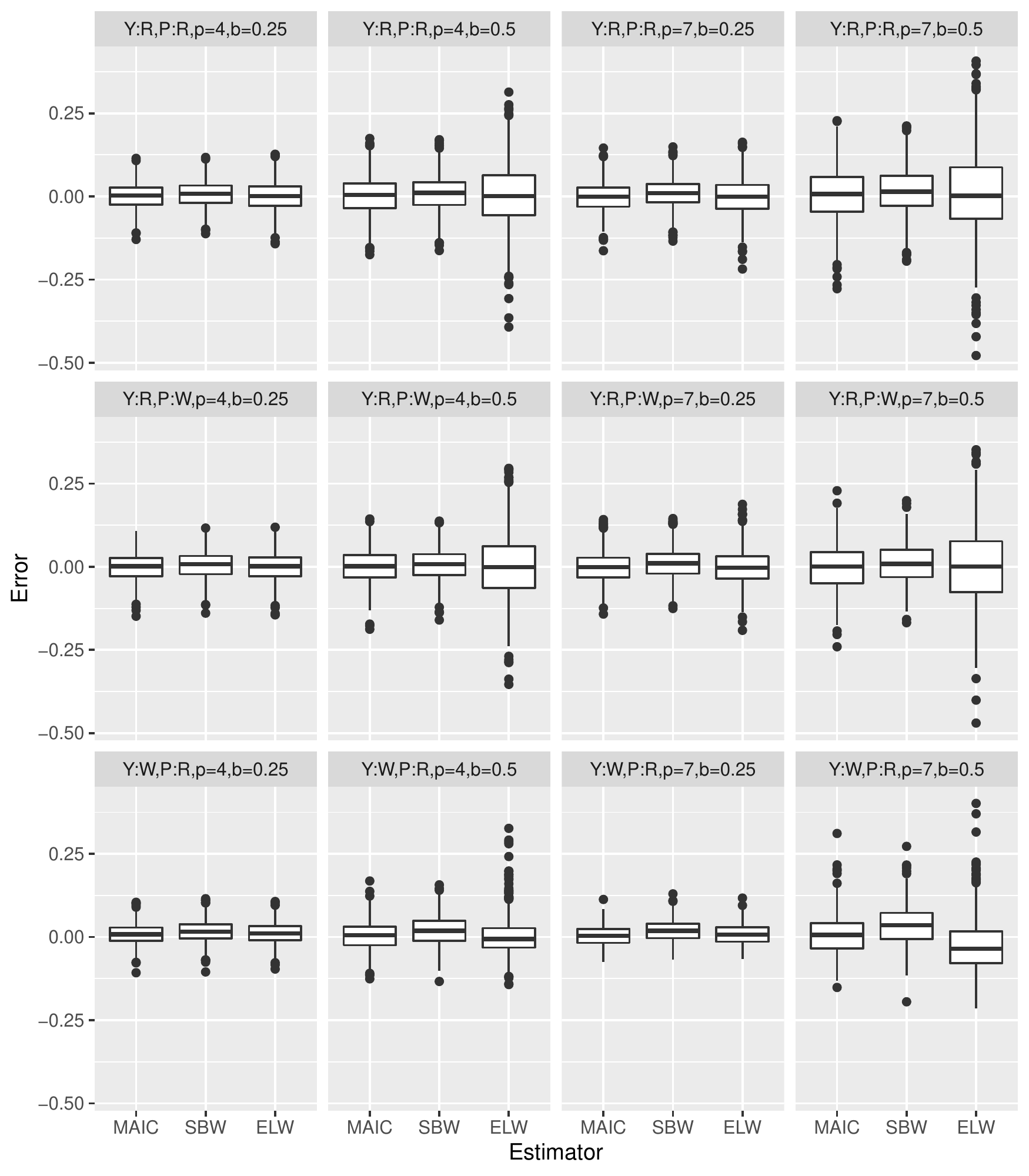}
    \caption{Error distribution of weighted estimators (\ref{weiest}) with MAIC weights (MAIC), stable balancing weight (SBW), and empirical likelihood weights (ELW), with $p=4, 7$ and $b=0.25, 0.5$,  when both models are correct (Y:R,P:R), only Y-model is correct (Y:R, P:W) and only P-model is correct (Y:W, P:R),  with 1000 simulation runs each. In the headers, Y and P are shorthand for Y-model and P-model, and R and W for Right and Wrong, respectively.}
    \label{fig1}
\end{figure}

\section{Discussion}
We have illustrated the connection between MAIC and CE, a relatively new area in survey sampling with a rich literature in recent years.  It is impossible to list all approaches and their properties that may be borrowed to improve indirect comparisons. A few of them are briefly described below, among discussions on other issues.  

One important advantage of using empirical likelihood weights is that the inference on the treatment effect can be based on the empirical likelihood with well-known properties \cite{owe01}. Based on the empirical likelihood ratio, one can construct a confidence interval for $\mu_1$, which is often more accurate than that based on the asymptotic normality required for MAIC and stable balancing weights.  

Variance estimation for MAIC and other CE estimators is a challenging task. Our simulation showed that the proposed SE estimator and the 95\% CI based on it generally behaved well when the sample size was large or the differences in mean covariates were small.  But they might be overly optimistic when the number of covariates to be balanced was large relative to the sample size or the covariate mean differences were large,  both making balancing more difficult.  As the coverage of 95\% CI may also be affected by non-normality of the MAIC and other CE estimators,  it is prudent to also use nonparametric bootstrap in situations with these potential issues. Based on our experience, although difficult in a simulation study, applying a nonparametric bootstrap with, eg, 1000 bootstrap runs for a specific analysis rarely causes numerically issues, unless balancing is difficult to achieve.  

We have only conducted a rather limited simulation study to compare the three calibration estimators. Although the results alone are not sufficient to determine which one may perform better, together with their theoretical properties as discussed in Section 3, they may shed some light on the direction for further investigation. For example, it seemed that both MAIC and stable weight estimators had less variability, while the latter may not be DR under the simulation settings here. Nevertheless, the latter may indeed have smaller variance. Therefore, assumptions on the Y- and P-models may have a potential impact on the performance of the estimators.  A more reliable approach to the choice of a method (if a choice has to be made) in an application is to perform a simulation under specific scenarios.  This can be done easily by either using off-the-shelf packages such as ebal \cite{ebal} and sbw \cite{zub20}, or adapting R packages for specific optimization problems.  The latter approach is particularly efficient for simulation or bootstrap and has more control over the optimization process.   

Another distance function not commonly used is the absolute dispersion $|w_i-1/n|$.  As shown in \cite{wan20}, it might lead to better performance in terms of MSE in some situations. Theoretical properties of minimizing $|w_i-1/n|$ weights have not been examined. Again, a simulation study under specific scenarios is a practical and reliable way to assess its performance in a specific scenario.  

One major issue in using CE for indirect comparison is the impact of extreme weights when the two populations are quite different and the number of covariates to be matched is not small, as shown in our simulation.  This is a common issue in any confounding adjustment. The IPW approach \cite{ros83} can mitigate the issue due to a large number of covariates, but may be less efficient than CE. One can also adapt CE approach by adding a lasso penalty in the optimization problem (\ref{gence}) \cite{tan20}.   The stable balancing weights estimator can also deal relatively a larger number of covariates, at the cost of imbalance in $\bX_i$ to some extent. The selection of $\bd$ is important to make a good compromise \cite{cha20}.  Another approach is to use model assisted CE that balances the predicted outcomes based on the Y-model \cite{wu01}.  This approach may not be feasible if the information from the target population is from publications.  However, in multi-party settings, the owner of the target population may be able to calculate the mean predicted outcome based on a fitted Y-model, if  collaboration between parties is feasible.  The approach can be made multirobust by balancing multiple Y-models and can be implemented by all the three CE approaches.  

The results of some recent simulations  and discussions around them \cite{pet18,hat20,phi20c, phi21,rem20,rem20b} rised interesting practical and methodological questions on the performance of MAIC and how to evaluate it.  
One interesting question is whether other CE approaches could perform significantly better than MAIC when it performed badly in some scenarios.  It seemed that several  scenarios in which MAIC had bad performance lackes overlap between the distributions of $\bX_{ij}$, or misspecified P- and Y- models.  Without the constraint of non-negative weights, the CE estimator with squared $D(w_i,1/n)=(w_i-1/n)^2$ may perform better with limited overlap when the Y-model is linear, since it is equivalent to extrapolation with a linear model.  With $w_i \ge 0$, our simulation results showed the squared distance weights performed similar to the MAIC weights.  
Nevertheless, the CE framework provides systematic approaches to improving its methods by using  appropriate distance function, balancing tolerance, and functions of $\bX_i$ to balance.    
Following the last one, model-assisted CE is a promising approach to deal with the issue with lack of overlap and nonlinear Y-modela, especially when the lack of overlap is due to a large $p$.  In summary, much further researches including carefully designed simulation studies are needed to modify and improve CE approaches including MAIC, and to evaluate them in relevant scenarios.  

{\bf Acknowledgment}
The author would like to thank an associate editor and  two referees for detailed comments and suggestions which led to  substantial improvement of the manuscript.

{\bf Data Availability Statement}
Data sharing not applicable to this article as no datasets were generated or analysed during the current study

\section{Appendix 1: On the implementation of weight calculation algorithms}
One key step in CE is weight calculation. Some software specialised for one of the three approaches, some also including weighted analyses such as sbw \cite{zub20} for the stable balancing weights, ebal \cite{ebal} for entropy balancing, and R-function Lag2 \cite{wu05} for empirical likelihood can be used.  Each of them used one optimization algorithm; ebal used R-function optim(), while sbw provides a few quadratic programming algorithms to choose. However, for simulation or bootstrapping, they might be inefficient and/or unstable. 

As the task can be formed as a constraint optimization problem, often one can use an optimization package to calculate the weights. This way gives more access to key parameters to control the algorithm.  For calculating entropy balancing weights, it is more efficient to minimize the dual form of (\ref{ebal}) \cite{hai11}:
\begin{equation*}
    \log(\sum_{i=1}^n \exp (-\bgamma^T \bX_i))+\bgamma^T \bar \bX_0.
\end{equation*}
  For stability and efficiency, we used the BFGS method, a quasi-Newtow approach \cite{fle87} implemented in R-function optim \cite{R}. The simulation showed that it had less numerical issues than sbw in the simulation study.  For SBW, an R package quadprog (one of the choices in package R-function sbw) was used directly for more efficiency and control. The R-function Lag2 \cite{wu05} is small but reasonably stable and was used directly, except with additional control of the maximum number of iterations to avoid infinite loops. R-codes for the implementation for a simulation are given in On-line Appendix.

\section{Appendix 2: R code for estimating $\hat V_{2s}$}
\baselineskip=7pt
\begin{verbatim}
simu=function(BB=50,
  nsimu=2000,
  nsub=500, #nsub=n_1
  n0=2000,
  np=3,
  b=0.5,
  bbeta=0.3){
  # function for MAIC
  fn=function(bb){
    log(sum(exp(-Xi0%*%bb)))+sum(bb*mp1)
  }
  # gradients
  grr=function(bb){  
    -t(Xi0)%*%exp(-Xi0%*%bb)/sum(exp(-Xi0%*%bb))+mp1
  }
    beta=rep(bbeta,np)
  set.seed(134)
  Xif=matrix(nrow=n0,rnorm(n0*np))
  mp1=apply(Xif,2,mean)
  Xm=apply(Xif,2,mean)
  Mx=matrix(rep(Xm,nsub),nrow=nsub,byrow=T)
  mu1=mean(Xif%*%beta)
  Out=NULL
  for (j in 1:nsimu){
    Xin=matrix(nrow=nsub,rnorm(nsub*np))+b
    Yi=Xin%*%beta +0.5*rnorm(nsub)
    Outb=NULL
    #Bootstrap runs
    for (j in 1:BB){
      ind=sample(1:nsub, replace=TRUE)
      Xi0=Xin[ind,]
      n=dim(Xi0)[1]     
      p=dim(Xi0)[2]
      opt=try(optim(par=rep(0,p),fn=fn,gr=grr,method="BFGS",
      control=list(maxit=300, reltol=1e-5))) #Find gamma for MAIC
      bb=opt$par
      wei=c(exp(-Xi0%*%bb)/sum(exp(-Xi0%*%bb))) #MAIC weights
      est=sum(Yi[ind]*wei) #MAIC estimator
      Outb=c(Outb,est)
    }
    Vbb=var(Outb,na.rm=TRUE) #Var by bootstrap
    Xi0=Xin
    opt=try(optim(par=rep(0,p),fn=fn,gr=grr,method="BFGS", 
    control=list(maxit=300, reltol=1e-8))) #Find gamma for MAIC
    bb=opt$par
    wei=c(exp(-Xi0%*%bb)/sum(exp(-Xi0%*%bb))) #MAIC weights
    est=sum(Yi*wei) #MAIC estimator
    Xim=Xi0-Mx
    A11=t(Xi0)%*%diag(wei)%*%Xi0- outer(Xm,Xm) +1e-6
    A12=t(Xi0)%*%((Yi-est)*wei)
    Ainv=solve(A11)
    S12=(Xi0-Xm)%*%t(Ainv)%*%A12
    Si=sum(wei^2*(Yi-est-S12)^2) #2S-Var
    B22=sum(wei^2*(Yi-est)^2)  #MAIC Var
    CV2s=1*(abs(mu1-est)<1.96*sqrt(Si))
    CVbb=1*(abs(mu1-est)<1.96*sqrt(Vbb))
    Out=rbind(Out,c(mean(Yi-est),est-mu1,CV2s,CVbb,Si,Vbb,B22))
  }
  junk=c(apply(Out,2,mean), var(Out[,2]))
  c(nsub,bbeta,b,np,junk[1:4],sqrt(junk[5:8]))
}

simu(nsimu=2000,nsub=100)
\end{verbatim}
\end{document}

%% file: newcom.tex
\newcommand{\balpha}{{\mbox {\boldmath$\alpha$}}}

\newcommand{\bbeta}{{\mbox {\boldmath$\beta$}}}

\newcommand{\bgamma}{{\mbox {\boldmath$\gamma$}}}

\newcommand{\epsi}{\varepsilon}

\newcommand{\bX}{\mbox {\bf X}}
\newcommand{\bZ}{\mbox {\bf Z}}

\newcommand{\bD}{\mbox {\bf D}}

\newcommand{\bI}{\mbox {\bf I}}

\newcommand{\bd}{\mbox {\bf d}}

\newcommand{\bm}{\mbox {\bf m}}

\newcommand{\bone}{\mbox {\bf 1}}

\newcommand{\var}{\mbox {var}}

\newcommand{\beqn}{\begin {equation}}
\newcommand{\eeqn}{\end {equation}}
\newcommand{\beqa}{\begin {eqnarray}}
\newcommand{\eeqa}{\end {eqnarray}}

\newcommand{\bvbt}{\begin{verbatim}}
\newcommand{\benu}{\begin{enumerate}}
\newcommand{\eenu}{\end{enumerate}}
\newcommand{\bver}{\begin{verbatim}}

\newcommand{\nn}{\nonumber}
\makeatletter
\input{rotate}\newbox\rotbox
{\par\vskip\z@\egroup \rotl\rotbox \end{table}}
\makeatother
\newdimen\x \x=1.5mm